\documentclass[aps, prl, reprint, superscriptaddress,preprintnumbers,amssymb]{revtex4-2}
\usepackage{graphicx}
\usepackage{dcolumn}
\usepackage{bm}

\begin{document}

\preprint{ver6}

\title{Altermagnetic spin-split Fermi surfaces in CrSb revealed by quantum oscillation measurements}

\author{Taichi Terashima}
\email{TERASHIMA.Taichi@nims.go.jp}
\affiliation{Research Center for Materials Nanoarchitectonics (MANA), National Institute for Materials Science, Tsukuba 305-0003, Japan}
\author{Yuya Hattori}
\altaffiliation[Present address: ]{Department of Physics, Tokyo Metropolitan University, 1-1, Minami-osawa, Hachioji 192-0397, Japan}
\affiliation{Center for Basic Research on Materials (CBRM), National Institute for Materials Science, Tsukuba 305-0047, Japan}
\author{David Graf}
\affiliation{National High Magnetic Field Laboratory, Florida State University, Tallahassee, Florida 32310, USA}
\author{Takahiro Urata}
\email{urata.takahiro.k4@f.gifu-u.ac.jp}
\altaffiliation[Present address: ]{Department of Electrical, Electronic and Computer Engineering, Gifu University, Gifu, Gifu 501-1193, Japan}
\affiliation{Department of Materials Physics, Nagoya University, Chikusa-ku, Nagoya 464-8603, Japan}
\author{Tomoki Yoshioka}
\affiliation{Department of Materials Physics, Nagoya University, Chikusa-ku, Nagoya 464-8603, Japan}
\author{Wataru Hattori}
\affiliation{Department of Materials Physics, Nagoya University, Chikusa-ku, Nagoya 464-8603, Japan}
\author{Hiroshi Ikuta}
\affiliation{Department of Materials Physics, Nagoya University, Chikusa-ku, Nagoya 464-8603, Japan}
\affiliation{Research Center for Crystalline Materials Engineering, Nagoya University, Chikusa-ku, Nagoya 464-8603, Japan}
\author{Hiroaki Ikeda}
\email{ikedah@fc.ritsumei.ac.jp}
\affiliation{Department of Physics, Ritsumeikan University, Kusatsu, Shiga 525-8577, Japan}

\date{\today}

\begin{abstract}
We report a comprehensive quantum oscillation study of the prototypical altermagnet CrSb, combining high-field magnetotransport and torque measurements with DFT + $U$ calculations including spin-orbit coupling. 
Multiple quantum oscillation frequencies were observed and tracked over wide angular ranges. 
The measured frequency branches are consistently explained by the spin-split Fermi surfaces arising from the altermagnetic electronic structure.
Our determined Fermi surface reveals that bands 1 and 2 form closed pockets centered at the A point, rather than the tubular $c$-axis-open sheets or $\Gamma$-centered closed pockets proposed in previous studies.
Our findings establish the Fermi-surface topology of CrSb and provide a firm basis for exploring emergent phenomena in altermagnetic materials.
\end{abstract}

\maketitle

In nonmagnetic materials with a centrosymmetric crystal structure, the electronic bands are doubly or even-fold degenerate due to the spin degree of freedom.
Ferromagnetic ordering lifts the spin degeneracy, splitting the up- and down-spin bands.
In contrast, antiferromagnetic ordering is usually thought to preserve the spin degeneracy:
strictly, the spin degeneracy remains if the product of the space inversion (plus a lattice translation if necessary) and the time reversal is a symmetry of the ordered state (see, for example, ref.\onlinecite{Cvetkovic13PRB}).

Recently, however, spin splitting and its consequences in some types of antiferromagnets have attracted much attention,
partially because they might confer superior functionalities in spintronic applications.
One noteworthy observation is the spontaneous anomalous Hall effect, which is usually associated with ferromagnetism, in Mn$_3$Sn \cite{Nakatsuji15Nature} and hexagonal FeS \cite{Takagi25NatMat}.
Both compounds are antiferromagnets (the former non-collinear, the latter collinear) with vanishingly small net magnetization.

We focus on the collinear case because CrSb is a collinear antiferromagnet. 
Early work on FeF$_2$ and MnO$_2$ \cite{Lopez-Moreno12PRB, Noda16PCCP} was followed by a flurry of theoretical studies on antiferromagnets with spin-split energy bands around 2020, including band-structure calculations, symmetry considerations, and calculations of spin current and the anomalous Hall effect \cite{Ahn19PRB, Naka19NatCommun, Hayami19JPSJ, Smejkal20SciAdv, Yuan20PRB, Mazin21PNAS}.
Subsequently, the term `altermagnetism' was coined for this type of antiferromagnetic ordering \cite{Smejkal22PRX} (see \cite{Mazin24Physics} for a brief historical explanation).
The up- and down-spin magnetic sublattices in altermagnets are connected by rotation, rather than by translation or inversion as in usual antiferromagnets.
The electronic band structures of altermagnets exhibit momentum-dependent spin splitting, which can be characterized as planar or bulk $d$-, $g$-, or $i$-waves.

RuO$_2$ has been identified as the prime metallic candidate for experimental verifications of altermagnetism.
Theoretically predicted as an altermagnet with $d$-wave symmetry \cite{Ahn19PRB, Smejkal20SciAdv}, RuO$_2$ has nonetheless been confirmed as nonmagnetic in recent $\mu$SR studies. 
The possible moment per Ru atom is only $\sim10^{-4}\mu_B$ \cite{Hiraishi24PRL, Kessler24npjSpin}.
Moreover, the frequencies obtained in quantum oscillation measurements on RuO$_2$ are consistent with the nonmagnetic state \cite{Wu25PRX}.
Another promising candidate is MnTe, which exhibits altermagnetic spin splitting as confirmed in angle-resolved photoemission spectroscopy (ARPES) measurements \cite{Krempasky24Nature, Lee24PRL, Osumi24PRB}.
However, as MnTe is a semiconductor, it is unsuitable for confirming spin splitting through quantum oscillation measurements. 

In contrast, CrSb is a metallic compound that
crystallizes as a hexagonal NiAs-type structure in the centrosymmetric space group P6$_3$/mmc (\#194) [inset of Fig. 1(a)].
The unit cell contains two Cr atoms connected by a 6-fold screw rotation (6$_3$).
Neutron diffraction measurements have firmly established that one Cr is an up-spin site while the other is a down-spin site \cite{Snow52PR, Takei63PR}.
The ordered moment in the ordered state below $T_N \sim$ 700 K is $\sim3\mu_B$ parallel to the $c$ axis, indicating an altermagnetic order with bulk $g$-wave symmetry.
Moreover, ARPES measurements have observed large $g$-wave spin-splitting of the electronic bands in CrSb \cite{Reimers24NatCommun, Yang25NatCommun, Zeng24AdvSci, Ding24PRL, LiW25PRB, LiC25CommunPhys}.
In this work, we confirm the bulk electronic band structure with $g$-wave spin splitting through quantum oscillation measurements.


\begin{figure*}
\includegraphics[width=15cm]{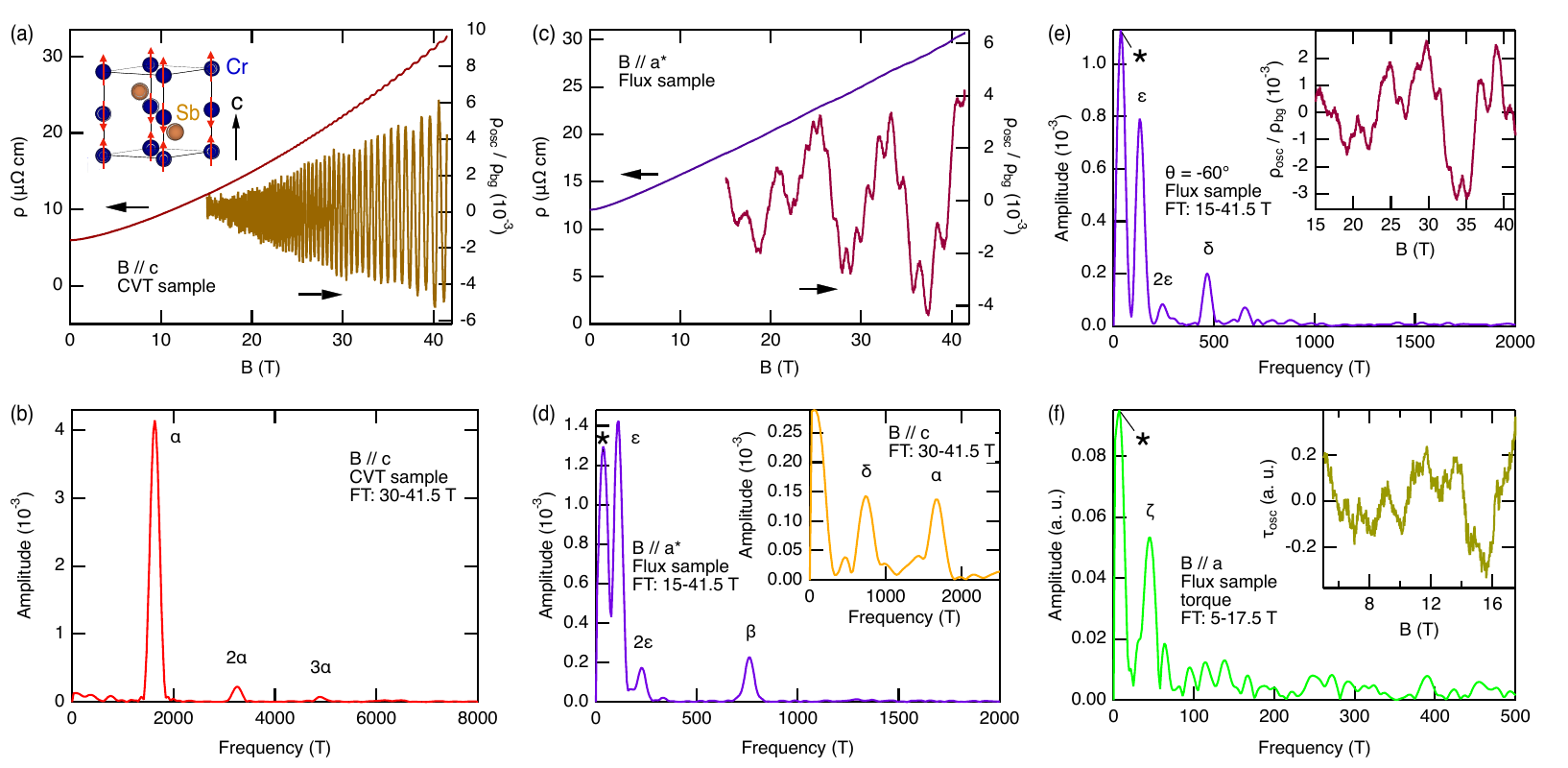}
\caption{\label{Signals}Quantum oscillations in CrSb. 
(a) Magnetic field dependence of the resistivity for $B \parallel c$ in the CVT-grown sample, along with the oscillatory component (SdH oscillations) $\rho_{osc}$ obtained by subtracting a polynomial background $\rho_{bg}$. 
The upper-left inset shows the crystal and magnetic structures of CrSb drawn in VESTA software \cite{Momma11JAC}.
(b) Corresponding Fourier transform. 
(c) Magnetic field dependence of the resistivity for $B \parallel a^*$ in the flux-grown sample, and the SdH oscillations obtained by subtracting a polynomial background. 
(d) Corresponding Fourier transform and (inset) Fourier transform for $B \parallel c$ in the same flux-grown sample.
(e) Fourier transform of SdH oscillations (inset) at field angle $\theta = -60^{\circ}$ in the flux-grown sample. $\theta$ is measured from the $c$ axis toward the $a^*$ axis. 
(f) Fourier transform of magnetic torque dHvA oscillations (inset) for $B \parallel a$. 
The $\alpha$, $\beta$, $\delta$, $\epsilon$, and $\zeta$ frequency peaks are labeled in (b), (d), (e), and (f). 
Harmonics of $\alpha$ and $\epsilon$ are also indicated.
The lowest frequency peaks marked with asterisks in (d), (e), and (f) are neglected (see text for explanation). 
}
\end{figure*}

CrSb single crystals were grown via two methods: the Sn-flux method and the chemical-vapor-transport (CVT) method \cite{Urata24PRM, Peng25PRB} (see Supplementary Material \cite{*[][ for the crystal growth procedure{,} Lifshitz-Kosevich formula{,} band-structure calculations{,} comments on {[*]}{,} Landau-fan plot of {$\alpha$} oscillations{,} and experimental and calculated Fermi surface parameters{,} which includes Refs{.} *--*] SM}).
The flux-grown crystals were hexagonal rods with axes parallel to the $c$ axis. The CVT-grown crystals were plate-shaped with surfaces perpendicular to the $c$ axis.
Accordingly, the electric currents for measuring the resistivities of the flux- and CVT-grown crystals were applied along the $c$ and $a$ axes, respectively.
Shubnikov--de Haas (SdH) oscillations were observed in two magnet/cryostat setups:
one using a 41.5-T resistive magnet and a He-3 refrigerator at the National High Magnetic Field Laboratory (NHMFL); the other using a 20-T superconducting magnet (up to 17.5 T) and a dilution refrigerator at the National Institute for Materials Science (NIMS).
Consistent results were obtained from multiple samples grown via each method.
This article focuses on our most thoroughly investigated samples: a flux-grown crystal with a residual resistivity ratio (RRR) of 6.8 and a CVT-grown crystal with an RRR of 9.8.

To observe Haas-van Alphen (dHvA) oscillations, we also measured magnetic torque on other flux-grown crystals using a micro cantilever at NIMS. 

We briefly clarify the quantum-oscillation frequency $F$.
$F$ is related to the maximum or minimum cross-sectional area $A$ of the Fermi surface normal to the applied field via the Onsager relation: $F = (\hbar/2\pi e) A$ \cite{SM}.
In the present case, $F$ = 300 T corresponds to 1\% of the cross-sectional area of the Brillouin zone perpendicular to the $c$ axis.
The cross-sectional area $A$ depends on where the Fermi surface is cut, which can be specified by the wave number $\kappa$ along the field direction. 
In this context, the adjectives `maximum' and `minimum' refer to the $A$'s at which d$A$/d$\kappa$ = 0.
The Fermi surface is determined by comparing experimental frequencies to theoretical predictions based on band-structure calculations.
The temperature and magnetic-field dependences of the quantum-oscillation amplitude are described by the Lifshitz--Kosevich formula \cite{SM}.

We performed fully relativistic electronic band-structure calculations including spin-orbit coupling (SOC) for the altermagnetic state, using the PBE-GGA potential within the DFT~(density functional theory)~+~$U$ scheme with the \textsc{Wien2k} code \cite{WIEN2K_JCP, Perdew96PRL} (see Supplemental Material \cite{SM}).
SOC is essential for the data interpretation as it alters the cyclotron orbits by gapping out the crossing points of Fermi surface sheets on the nodal planes $\Gamma$MK, $\Gamma$KHA, and ALH, where altermagnetic spin splitting does not occur (the high-symmetry points and lines in the Brillouin zone are explained in \cite{SM}).
In the following, the wave numbers $k_x$ ($\parallel a^*$) and $k_y$ ($k_x \bot k_y$) are measured in units of $(2/\sqrt{3})(2\pi/a)$, while $k_z$ is measured in units of $2\pi/c$.

\begin{figure*}
\includegraphics[width=17cm]{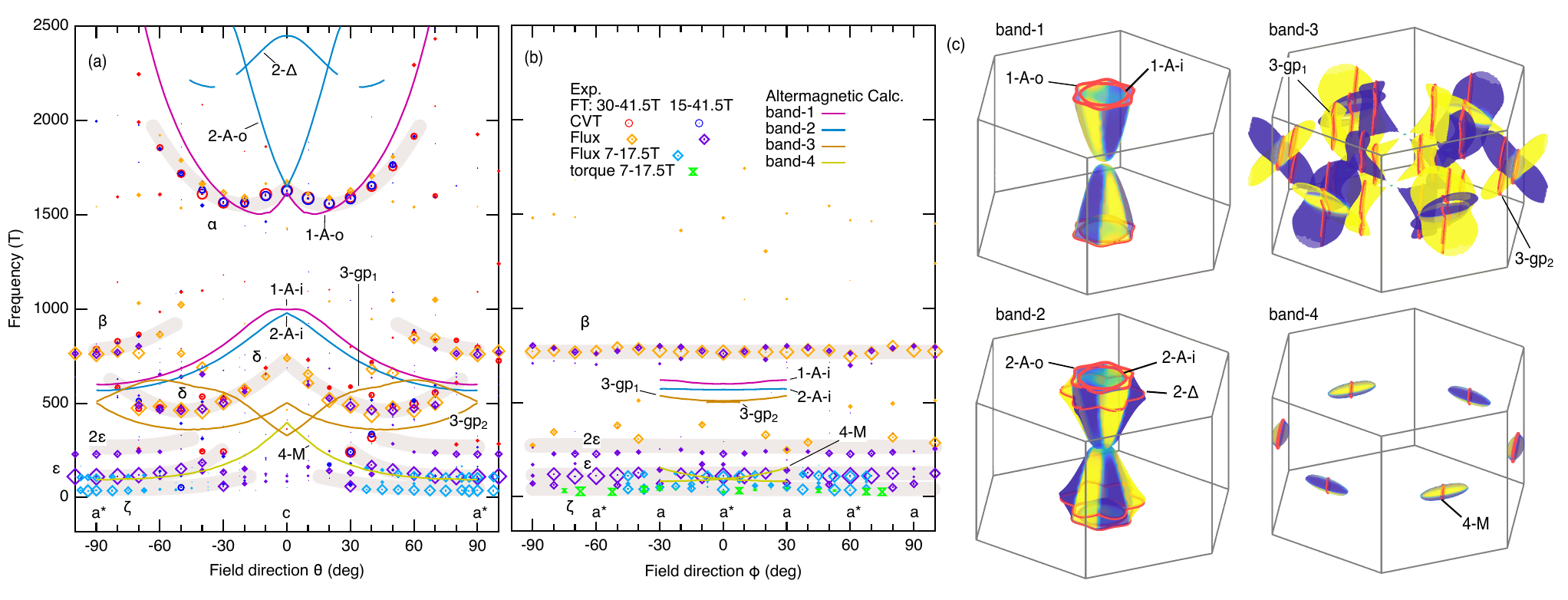}
\caption{\label{Signals}Altermagnetic Fermi surface in CrSb. 
(a) and (b) Magnetic-field-direction dependences of the quantum oscillation frequencies for field rotation in the (a) $ca^*$ and (b) $a^*a$ planes compared to altermagnetic calculations.
The experimental frequency branches are labeled with Greek letters, and the grey shading serves as an eye guide.
Different symbols correspond to different samples and/or field windows of Fourier transforms as indicated in (b), and their sizes indicate the oscillation amplitudes.
In (b), the 7–17.5 T SdH data for the flux-grown sample (light blue diamonds) cover the range from $\phi = -45^{\circ}$ to $+67.5^{\circ}$, and the torque measurements (green hourglass) cover the range from $-75^{\circ}$ to $+97.5^{\circ}$.
The solid curves plot the theoretical frequencies calculated from the altermagnetic Fermi surface in (c) for $-90 \leqslant \theta \leqslant 90^{\circ}$ (a) and $-30 \leqslant \phi \leqslant 30^{\circ}$ (b).
The attached names of the underlying orbits include the band number, orbit center, and, if necessary, inner (i) or outer (o).
gp denotes a general point.
(c) Altermagnetic Fermi surface obtained from our DFT+$U$ calculation (with SOC).
The color indicates the spin polarization $s_z$.
Extremal orbits relevant to experimental frequencies are indicated: $B \parallel c$ for bands-1 and 2, and $B \parallel a^*$ for bands-3 and 4.
}
\end{figure*}

Our final calculated Fermi surface is shown in Fig. 2(c). 
Four spin-non-degenerate bands, named bands 1, 2, 3, and 4 from lower to higher energy, cross the Fermi level [Fig. S1(b)] and form the respective Fermi surface sheets shown in Fig. 2(c).
In this calculation, we used $U = 0.03$ Ry, and to match the frequency $\alpha$ observed experimentally for $B \parallel c$ with the orbit 1-A-o on the band-1 Fermi surface (see below), we shifted the energies of bands 1 and 2 upward by 65 meV, and, to compensate for the resulting change in carrier number, shifted the energies of bands 3 and 4 downward by 9.56 meV.
Adjustment of the band energy by up to $\sim100$ meV is a common practice when comparing theoretical and experimental dHvA frequencies \cite{Carrington05PRB, Coldea08PRL, Analytis09PRL, Terashima11PRL, Baglo22PRL}.
Each of the band-1 and -2 Fermi surfaces consists of an outer rugby-ball-shaped pocket elongated along the $k_z$ direction, which is centered at the A point and truncated at the $\Gamma$ point, as well as an inner, ball-shaped pocket centered at the A point.
The band-3 Fermi surface consists of small pockets surrounding the $\Gamma$ point and X-shaped closed pockets at the M points (see also Fig. S2), while the band-4 surface consists of ellipsoidal pockets at the M points.
As indicated by the color coding, the spin orientation on each Fermi surface varies with momentum $k$, following $g$-wave symmetry. 
The Fermi sheets of bands 1 and 2, as well as those of bands 3 and 4, can be regarded as pairs that are altermagnetically spin split.

We now show selected quantum-oscillation data in Fig. 1.
SdH oscillations in both the flux- and CVT-grown samples [Figs. 1(a) and 1(c)] are already discernible in the raw resistivity versus field curves and become more evident after subtracting the polynomial backgrounds.
The frequencies $\alpha$, $\beta$, $\delta$, $\epsilon$, and $\zeta$ are labeled in the Fourier transform images [Figs. 1(b), 1(d), 1(e), and 1(f)].
The $\zeta$ oscillation was most visible in the magnetic torque data [Fig. 1(f)].
The lowest frequency peaks, marked with asterisks in Figs. 1(d), 1(e), and 1(f), are neglected because they correspond to oscillations with periods of 1.5 or lower within the measured field range. 
Therefore, they are likely to be artifacts caused by removal of the DC component during data processing, which results in zero amplitude at zero frequency, although the peaks in Figs. 1(d) and 1(e) may include contributions from the $\zeta$ oscillation.
The frequency $\alpha$ corresponds to $F_3$ in \cite{Du26SciChinaPhys}, $f_1$ in \cite{Long26condmat}, and $F_1$ in \cite{Thadathil26condmat}, while $\delta$ corresponds to $F_2$ in \cite{Du26SciChinaPhys}.
The existence of $F_1$ in \cite{Du26SciChinaPhys}, which would yield a single oscillation period in the widest field window (24--35 T) of \cite{Du26SciChinaPhys}, is questionable.

Figures 2(a) and 2(b) show the field-direction dependences of the quantum-oscillation frequencies for field rotations in the $ca^*$ and $a^*a$ planes, respectively.
In general, Fourier-transforming a high-field region (i.e., 30--41.5 T) emphasizes the high frequencies whereas Fourier-transforming a wide inverse field ($1/B$) region (i.e., 7--17.5 or 15--41.5 T) enhances the frequency resolution and better resolves the low frequencies.
Figures 2(a) and 2(b) also plot the frequencies expected from the altermagnetic Fermi surface in Fig. 2(c). 
For bands-3 and 4, we present only the frequencies relevant to the interpretation of the experimental data (all calculated frequencies are reported in Fig. S4).

\begin{figure}
\includegraphics[width=8.5cm]{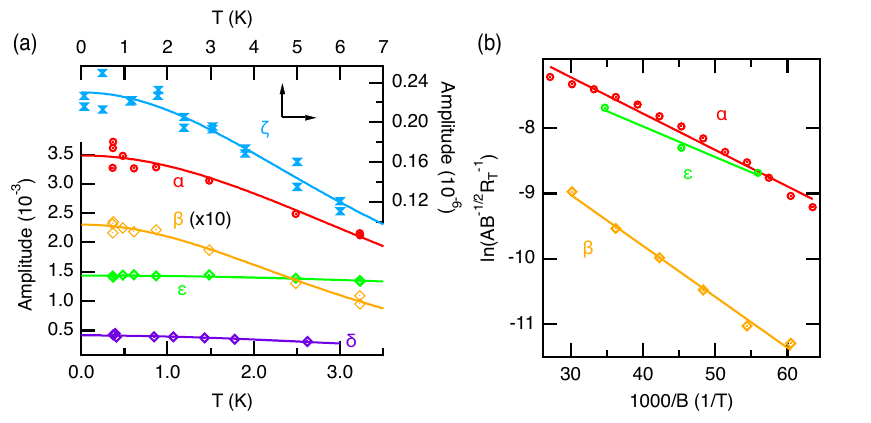}
\caption{\label{Signals}Temperature and magnetic-field dependences of quantum-oscillation amplitudes in CrSb. 
(a) Temperature dependences of the oscillation amplitudes of $\alpha$ for $B \parallel c$, $\beta$ and $\epsilon$ for $B \parallel a^*$, $\delta$ at $\theta = -60^{\circ}$, and $\zeta$ for $B \parallel a$.
The solid curves are the Lifshitz--Kosevich fittings, from which the effective masses are determined.
(b) Magnetic-field dependences of the oscillation amplitudes of $\alpha$ for $B \parallel c$ and $\beta$ and $\epsilon$ for $B \parallel a^*$.
The solid lines are linear fittings, from which the electron scattering times are estimated.
}
\end{figure}

Figure 3(a) shows the temperature dependences of the oscillation amplitudes along with the Lifshitz--Kosevich fittings \cite{SM}, from which the effective masses $m^*$ were determined as follows: $m^*/m_e$ = 1.2(1) for $\alpha$ with $B \parallel c$,  1.13(4), 0.29(3) for $\beta$ and $\epsilon$, respectively, with $B \parallel a^*$, 0.24(1) for $\zeta$ with $B \parallel a$, and 1.2(2) for $\delta$ at $\theta = -60^{\circ}$.
Here $m_e$ denotes the free electron mass, and $\theta$ is the angle between the applied field and the $c$ axis. 

As the field is tilted from the $c$ axis [Fig. 2(a)], the $\alpha$ frequency initially decreases and thereafter increases. 
The increase is observed at $|\theta| > 20^{\circ}$.
This angular dependence can be explained by the maximum orbit at the A point of the band-1 outer Fermi sheet [orbit 1-A-o in Fig. 2(c)].
In field directions near $B \parallel c$, the frequency from the minimum orbit at A of the band-2 outer sheet (orbit 2-A-o) approaches the frequency of orbit 1-A-o; hence, the $\alpha$ oscillation includes some contribution from orbit 2-A-o.
The effective mass of $\alpha$ is also in good agreement with band masses of these orbits (Table SI).
The band-2 outer sheet also has a maximum orbit at $k_z = 0.40$ (orbit 2-$\Delta$).
However, we find no experimental data points that clearly correspond to this orbit.


The $\beta$ frequency is not observed near $B \parallel c$, but becomes visible when the field is tilted away from the $c$ axis and is clearly observed for $|\theta| \gtrsim 60^\circ$.
This behavior can be explained by orbits 1-A-i and 2-A-i on the pockets formed by bands 1 and 2 at the $A$ point [Fig.~2(c)]. 
The band masses of these orbits begin to increase rapidly below $|\theta| \sim 50^\circ$ as the field direction approaches $B \parallel c$ from $B \parallel a^*$, and become approximately three times larger at $B \parallel c$ than at $B \parallel a^*$.
This naturally explains why the $\beta$ frequency is not observed near $B \parallel c$. 
In addition, the frequencies of these orbits exhibit only a very weak angular dependence within the $aa^*$ plane, which is also consistent with the behavior of the $\beta$ frequency [Fig. 2(b)].
Although the calculated frequencies are larger than the experimental $\beta$ frequency by approximately 200~T, this discrepancy corresponds to a band-energy shift of at most about 80~meV within the effective-mass approximation and is therefore well within the typical uncertainty of band-structure calculations. 

On the other hand, the experimental effective mass (1.13$m_e$) of $\beta$ for $B \parallel a^*$ is more than three times larger the band masses of 1-A-i and 2-A-i (Table SI). 
This observation suggests an alternative possibility. 
Namely, the $\beta$ oscillation may receive a contribution from orbit 3-gp$_1$ on the band-3 pocket near $\bm{k}_{\mathrm{gp}_{1}}$ =(0.27, 0, $-$0.23) (gp stands for general point). 
The frequency of 3-gp$_1$ is nearly identical to those of 1-A-i and 2-A-i, while its band mass is 0.90$m_e$ (Table SI). 
Therefore, the observed $\beta$ oscillation may contain contributions not only from 1-A-i and 2-A-i but also from 3-gp$_1$.

The $\delta$ frequency decreases as $|\theta|$ increases to $40^\circ$, and thereafter begins to increase. 
This behavior can be explained by an orbit 3-gp$_2$ centered at $\bm{k}_{\mathrm{gp_{2}}}$ = (0.44, 0, -0.09) on the X-shaped pocket at the M point of band 3 [Fig. 2(c)].
The effective mass determined at $\theta = -60^{\circ}$ is also consistent with the corresponding band mass (Table SI).
The discrepancy in frequency between experiment and calculation is at most about 200 T, but the corresponding difference in band energy is only less than 20 meV.
The $\epsilon$ frequency can be assigned to an orbit 4-M on the pocket at the $M$ point of band 4 [Fig. 2(c)]. 
The effective and band masses reasonably agree (Table SI).
$\epsilon$ is not observed near $B \parallel c$, which is presumably because the band mass of orbit 4-M increases as the field direction approaches $B \parallel c$. 
The $\zeta$ frequency is likely due to a small pocket that could not be reproduced by the calculations for band 3.

For the sake of completeness, we performed band-structure calculations for the nonmagnetic state (Fig. S3).
In this case, bands 1 and 2, and bands 3 and 4 are degenerate because of spin degeneracy.
The calculated Fermi surface is shown in Fig. S3(b), which is very different from the altermagnetic Fermi surface in Fig. 2(c).
We compared the experimental frequencies with the frequencies expected from the nonmagnetic Fermi surface [Fig. S3(c)].
The agreement is poor, and especially, the frequency branches $\alpha$ and $\delta$ can never be explained by the nonmagnetic calculation.

Thus, it has become clear that the altermagnetic Fermi surface shown in Fig. 2(c) matches well with the experimental results; let us now compare this Fermi surface with those reported in previous studies.

In the absence of both $U$ and band-energy adjustments, DFT calculations yield tube-like Fermi surfaces for bands 1 and 2 that are open along the $c$ axis [Fig. 4(a)]. 
Some ARPES studies have observed hole bands crossing the Fermi level near both the $\Gamma$ point and the $A$ point, which is consistent with the tube-like Fermi surfaces \cite{Yang25NatCommun, Zeng24AdvSci, LiW25PRB}. 
On the other hand, Long \textit{et al}. have observed quantum oscillations for $B \perp c$ and claimed that the Fermi surfaces of bands 1 and 2 are closed, centered at the $\Gamma$ point and terminated at the $A$ plane ($k_z=0.5$) \cite{Long26condmat}.

\begin{figure}
\includegraphics[width=8.5cm]{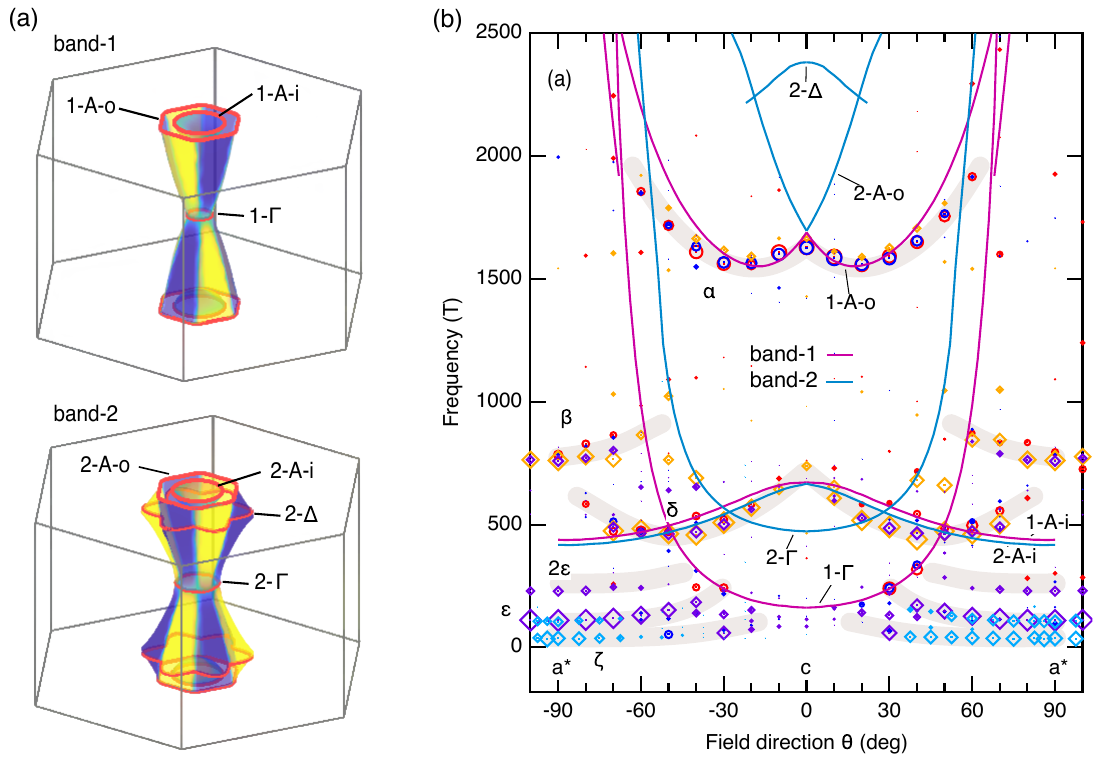}
\caption{\label{band}(a) Tube-like Fermi surfaces of bands 1 and 2 obtained from altermagnetic calculations without $U$.
(b) Calculated frequencies for the Fermi surfaces in (a) compared to experimental frequencies.
}
\end{figure}

Initially, we attempted to interpret our experimental results based on DFT calculations without $U$. 
In that case, the tube-like Fermi surfaces of bands 1 and 2 possess orbits 1-$\Gamma$ and 2-$\Gamma$ centered at the $\Gamma$ point [Fig. 4(a)]. 
Given that orbit 1-A-o at the A point produces prominent SdH oscillations, orbits 1-$\Gamma$ and 2-$\Gamma$ should likewise be detectable, particularly for $B \parallel c$ [Fig. 4(b)]. 
This expectation is further supported by their small calculated band masses, which are only about one third and two thirds of that of 1-A-o, respectively.
However, no frequencies corresponding to these orbits were detected.
We therefore performed calculations while varying $U$, and found that with $U = 0.03$ Ry and appropriate adjustments of band energies as explained above, the Fermi surfaces of bands 1 and 2 evolve into closed pockets centered at A and terminated at the $\Gamma$ plane ($k_z = 0$), as shown in Fig. 2(c). 
This provides an explanation for the absence of frequencies corresponding to 1-$\Gamma$ and 2-$\Gamma$ near $B \parallel c$.
Furthermore, orbit 1-A-o of Fig. 2(c) reasonably explains the $f_1$ frequency observed throughout the $ca^*$ plane, from $B \parallel c$ to $B \parallel a^*$, reported by Long \textit{et al.} \cite{Long26condmat} (see Supplemental Material \cite{SM}).
The tube-like Fermi surfaces at $U = 0$ are thicker at the A plane and thinner at the $\Gamma$ plane [Fig. 4(a)]. 
Therefore, our Fermi surfaces [Fig. 2(c)], which are terminated at the $\Gamma$ plane, provide a more suitable description than the A-plane-terminated Fermi surfaces proposed by Long \textit{et al.} \cite{Long26condmat}.
The observation of hole bands near the $\Gamma$ point in ARPES may be attributed to a slight shift of the band energies. 
Since our results place these bands only slightly below the Fermi level at $\Gamma$ [Fig. S1(b)], even a modest upward shift---for example, due to surface band bending---could render them observable in ARPES.

Figure 3(b) shows the Dingle plots indicating the field dependences of the $\alpha$, $\beta$, and $\epsilon$ oscillations, along with their linear fittings.
Using the Lifshitz--Kosevich formula \cite{SM}, the electron scattering times of the $\alpha$, $\beta$, and $\epsilon$ oscillations were estimated from the slopes as 0.38(4), 0.28(2), and 0.11(3) ps, respectively, and
the mobilities were estimated as 560(30), 400(20), and 700(100) cm$^2$/(Vs), respectively. The mobilities exceed those reported in \cite{Du26SciChinaPhys}
and approximate those estimated from multicarrier analyses of the magnetotransport data of major carriers with carrier densities of 10$^{20}$ cm$^{-3}$ or larger \cite{Urata24PRM, Peng25PRB}.
These multicarrier analyses suggested the existence of high-mobility minor carriers with mobilities above 2000 cm$^2$/(Vs) and carrier densities of $\sim$10$^{19}$ cm$^{-3}$ or lower, but such carriers could not be confirmed from the present SdH measurements.

Du \textit{et al}. claimed that the $\alpha$ oscillation ($F_3$ in \cite{Du26SciChinaPhys}) carries a nontrivial Berry phase $\pi$ \cite{Du26SciChinaPhys}.
However, the oscillation phase that we determined from a Landau-fan plot (Fig. S5) does not agree with that reported in \cite{Du26SciChinaPhys}.
In fact, the Berry phase cannot be definitively inferred from a phase analysis of the $\alpha$ oscillation because the $\alpha$ oscillation near $B \parallel c$ is the sum of the oscillations from the two orbits (see Supplemental Material \cite{SM}).

A recent theoretical study based on a two-dimensional model Hamiltonian predicted characteristic quantum-oscillation behavior in altermagnets, such as the splitting of a single frequency or the merging and re-splitting of two frequencies with increasing Zeeman field \cite{LiZ25PRB}. 
However, we were unable to observe such behavior in the present study.

In summary, quantum oscillation measurements combined with DFT+$U$ calculations reveal spin-split Fermi surfaces in the altermagnet CrSb. 
The observed frequency branches are consistently explained by an altermagnetic electronic structure, whereas nonmagnetic calculations fail to reproduce the data. 
Our results indicate that bands 1 and 2 form closed Fermi surfaces centered at the $A$ point rather than the previously proposed tubular or $\Gamma$-centered topologies. 
These findings establish the Fermi-surface topology of CrSb and provide a basis for future studies of altermagnetic phenomena.

\begin{acknowledgments}
This work was supported by JSPS KAKENHI Grant Numbers JP22K03537, JP23K25827, JP24KJ0227, and JP25K07202.
MANA is supported by the World Premier International Research Center Initiative (WPI), MEXT, Japan.
The National High Magnetic Field Laboratory is supported by National Science Foundation through NSF/DMR-2128556 and the State of Florida.
TU was supported by Foundation of Public Interest of Tatematsu.
We thank Troy Brumm for his valuable technical support at NHMFL.
TT thanks Hisatomo Harima for valuable discussion.
\end{acknowledgments}

%

\end{document}


\preprint{verSM5}

\title{Supplemental Material: Altermagnetic spin-split Fermi surfaces in CrSb revealed by quantum oscillation measurements}

\author{Taichi Terashima}
\email{TERASHIMA.Taichi@nims.go.jp}
\affiliation{Research Center for Materials Nanoarchitectonics (MANA), National Institute for Materials Science, Tsukuba 305-0003, Japan}
\author{Yuya Hattori}
\altaffiliation[Present address: ]{Department of Physics, Tokyo Metropolitan University, 1-1, Minami-osawa, Hachioji 192-0397, Japan}
\affiliation{Center for Basic Research on Materials (CBRM), National Institute for Materials Science, Tsukuba 305-0047, Japan}
\author{David Graf}
\affiliation{National High Magnetic Field Laboratory, Florida State University, Tallahassee, Florida 32310, USA}
\author{Takahiro Urata}
\email{urata.takahiro.k4@f.gifu-u.ac.jp}
\altaffiliation[Present address: ]{Department of Electrical, Electronic and Computer Engineering, Gifu University, Gifu, Gifu 501-1193, Japan}
\affiliation{Department of Materials Physics, Nagoya University, Chikusa-ku, Nagoya 464-8603, Japan}
\author{Tomoki Yoshioka}
\affiliation{Department of Materials Physics, Nagoya University, Chikusa-ku, Nagoya 464-8603, Japan}
\author{Wataru Hattori}
\affiliation{Department of Materials Physics, Nagoya University, Chikusa-ku, Nagoya 464-8603, Japan}
\author{Hiroshi Ikuta}
\affiliation{Department of Materials Physics, Nagoya University, Chikusa-ku, Nagoya 464-8603, Japan}
\affiliation{Research Center for Crystalline Materials Engineering, Nagoya University, Chikusa-ku, Nagoya 464-8603, Japan}
\author{Hiroaki Ikeda}
\email{ikedah@fc.ritsumei.ac.jp}
\affiliation{Department of Physics, Ritsumeikan University, Kusatsu, Shiga 525-8577, Japan}

\date{\today}

\maketitle

\renewcommand{\figurename}{FIG. S}
\renewcommand{\tablename}{TABLE S}

\renewcommand{\bibnumfmt}[1]{[SM#1]} 
\renewcommand{\citenumfont}[1]{SM#1}

\section{Crystal growth}
Single crystals of CrSb were grown using the Sn-flux and chemical vapor transport (CVT) methods.
The Sn-flux method (see \cite{Urata24PRM} for details)
yielded hexagonal-rod-shaped crystals. 
Out-of-plane x-ray diffraction and transmission Laue measurements revealed that the long axis was parallel to the $c$ axis, while the lateral facet was parallel to the $\{01\bar{1}0\}$ plane. 
The CVT method, performed as described in \cite{Peng25PRB}, 
yielded plate-like crystals with an edge angle of 120 degrees. 
Out-of-plane X-ray diffraction measurements of these samples confirmed that the largest facet was parallel to the (0001) plane. 
The in-plane crystal orientation was determined from transmission Laue measurements.

\section{Lifshitz--Kosevich formula for describing Shubnikov--de Haas oscillations}
A Shubnikov--de Haas oscillation $\rho_{osc}$ from an extremal orbit without spin degeneracy is expressed as \cite{Richards73PRB, Shoenberg84}
\begin{equation}
\frac{\rho_{osc}}{\rho_0} = -C\sqrt{B}R_TR_D \cos\left[2\pi\left(\frac{F}{B}-\frac{1}{2}\right)+\phi_D+\phi_B+\phi_Z\right], \tag{S1}
\end{equation}
where $\rho_0$ is the background resistivity, and $C$ is a positive coefficient. Harmonics are neglected.
We tacitly assumed that $\sigma = \rho/(\rho^2+\rho_H^2) \approx \rho^{-1}$, which is valid when the Hall resistivity $\rho_H$ is sufficiently smaller than $\rho$.
The frequency $F$ is related to the Fermi-surface cross-sectional area $A$ as $F = (\hbar/2\pi e)A$.
The temperature and Dingle reduction factors are given by $R_T= X/\sinh X$ and $R_D=\exp(-X_D)$, respectively, where $X_{(D)}=K\mu^* T_{(D)}/B$, $\mu^*=m^*/m_e$, and the coefficient $K$ is 14.69 T/K.

We define three phase offsets as follows:
$\phi_D$ is $+\pi/4$ and $-\pi/4$ when the oscillation arises from a minimum and maximum cross section, respectively.
$\phi_B$ is the Berry phase, which is 0 for normal electrons and $\pi$ for Dirac fermions \cite{Mikitik98JETP, Alexandradinata23ARCMP}.
$\phi_Z$ describes the phase shift due to the Zeeman energy. 
For a quadratic band dispersion, $\phi_Z$ is given by $\pi \bar{g} \mu^* / 2$, where $\bar{g}$ is an effective $g$ factor averaged over the cyclotron orbit \cite{Shoenberg84, Aoki14JPSJ}. For a linear band dispersion, $\phi_Z$ is not constant but depends on the magnetic field.

By fitting the temperature reduction factor $R_T$ to the experimental temperature dependence of the oscillation amplitude, one can estimate the effective mass $m^*$. 
The Dingle temperature $T_D$ is estimated from a plot of $\ln[A/(\sqrt{B}R_T)]$ against $1/B$, where $A$ is the oscillation amplitude. Such a plot is called a Dingle plot \cite{Shoenberg84}.
According to the form of $R_D$, the Dingle plot should be linear with a slope of $-K\mu^*T_D$.
The Dingle temperature $T_D$ is inversely proportional to the carrier scattering time $\tau$: specifically, $T_D=\hbar / 2\pi k_B\tau$.
Moreover, the mobility can be estimated as $\mu$ = $e\tau/m^*$. 

\section{Band structure calculations}

Fully relativistic spin-polarized electronic-structure calculations including spin-orbit coupling (SOC) were performed using the \textsc{Wien2k} code~\cite{WIEN2K_JCP}.
The exchange-correlation potential was treated within the generalized gradient approximation in the Perdew-Burke-Ernzerhof (PBE) form~\cite{Perdew96PRL}. 
The electronic structure was converged using a $7\times 7\times 5$ $k$-point mesh with RKmax $=7$, and the crystallographic parameters were taken from experimental data~\cite{Takei63PR}.
Space group $\#194$ was used for the nonmagnetic state, while space group $\#164$ was used for the altermagnetic state.
The Brillouin zone and high-symmetry points are shown in Fig. S1(a).

In the altermagnetic state, electron correlation effects were treated within the DFT+$U$ scheme~\cite{Liechtenstein95PRB, Shick99PRB, Petukhov03PRB}  employing the around-mean-field (AMF) double-counting correction, with the Hubbard $U$ applied to the Cr $d$ orbitals.
A Wannier-based tight-binding Hamiltonian was constructed using the \textsc{Wannier90} code~\cite{Marzari97PRB} interfaced with \textsc{Wien2k} via \textsc{Wien2kwannier}~\cite{Kunes10CPC}.
Based on this Hamiltonian, the Fermi surfaces were constructed, and the quantum oscillation frequencies and cyclotron effective masses were evaluated.

In the altermagnetic state,  four bands cross the Fermi level.
Without $U$, bands 1 and  2 form tube-like hole surfaces enclosing the $\Gamma-A$ line and closed pockets at the A point [Fig. 4(a)].
Band 3 forms a complex electron surface, while band 4 consists of cigar-shaped electron pockets located at the Brillouin-zone boundary.

As reported in Ref.[\onlinecite{LiC25CommunPhys}], the band-3 Fermi surface undergoes a significant reconstruction upon the introduction of a small Hubbard $U$.
The inclusion of $U$ = 0.03 Ry enlarges the volumes of the band-1 and -2 Fermi surfaces and breaks the complex connectivity of the band-3 Fermi surface, generating several small pockets.

To reproduce the experimentally observed oscillation frequency $\alpha$ with the extremal orbit 1-A-o of the band-1 Fermi surface obtained in the DFT+$U$ calculation, the Fermi levels of bands 1 and 2 were shifted upward by $0.065$ eV.
For bands 3 and 4, the Fermi level was shifted downward by $0.00956$ eV to compensate for the associated change in carrier number, which is $\sim$0.0204 per unit cell.
Adjustment of the band energy by up to $\sim100$ meV is a common practice when comparing theoretical and experimental dHvA frequencies \cite{Carrington05PRB, Coldea08PRL, Analytis09PRL, Terashima11PRL, Baglo22PRL}.

Figure S1(b) shows the calculated band structure.
Although the calculated Fermi surface is presented in Fig. 2(c), the band-3 Fermi surface is complicated and hence we show its top and side views in Fig. S2.

\begin{figure}
\includegraphics[width=15cm]{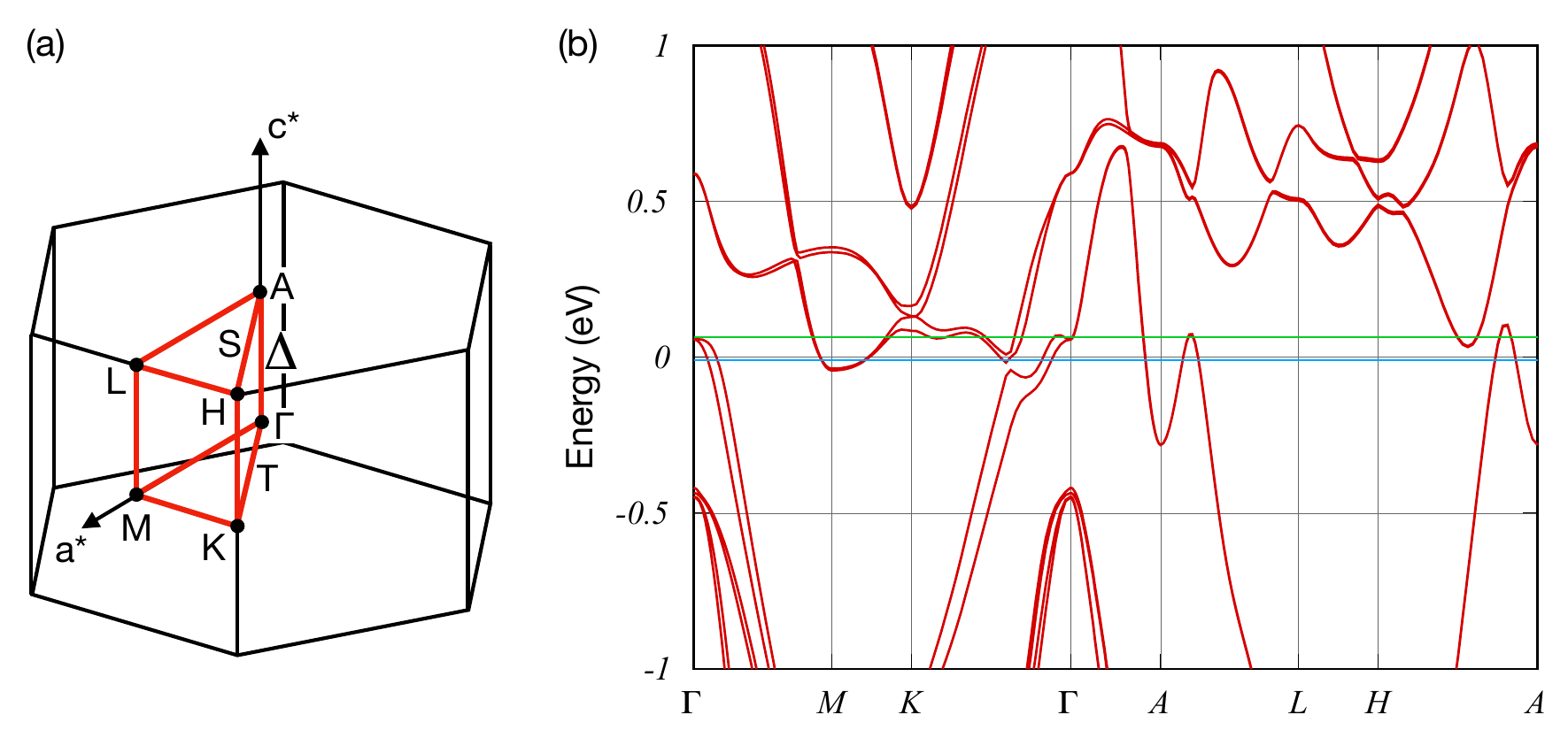}
\caption{\label{band}(a) Brillouin zone and high-symmetry points.
The high-symmetry lines $\Delta$, S, and T are also indicated.
The $\Gamma$MK, $\Gamma$KHA, and ALH planes are nodal planes of altermagnetic spin splitting without spin-orbit coupling.
(b) Calculated band structure of CrSb in the altermagnetic state with $U$ = 0.03 Ry.
The original Fermi level is at $E$ = 0.
The green and light blue horizontal lines show the shifted Fermi level for bands 1 and 2 and that for bands 3 and 4, respectively.
}
\end{figure}

\begin{figure}
\includegraphics[width=15cm]{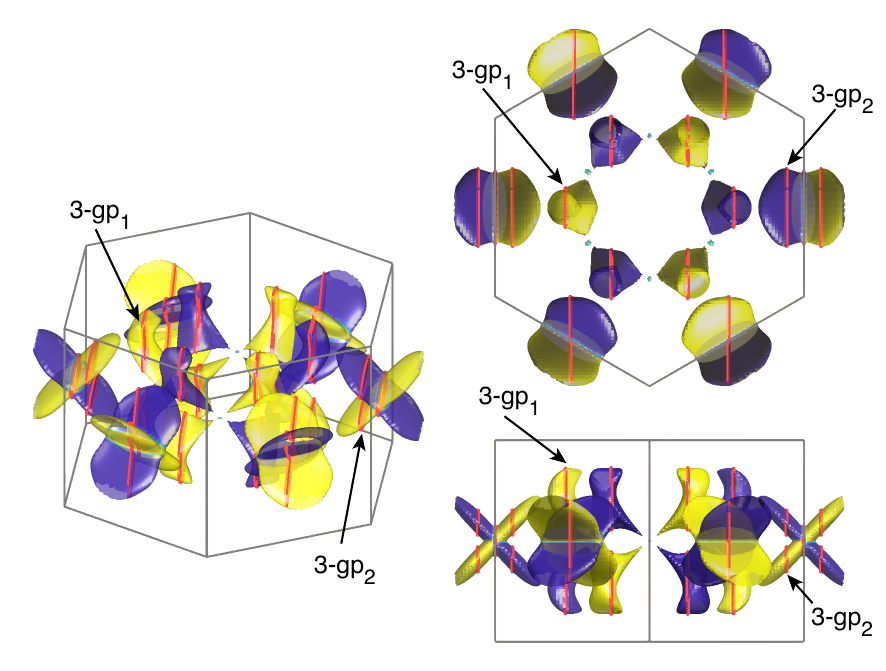}
\caption{\label{band} Band-3 Fermi surface of the altermagnetic state with $U$ = 0.03 Ry.
}
\end{figure}

\begin{figure}
\includegraphics[width=17cm]{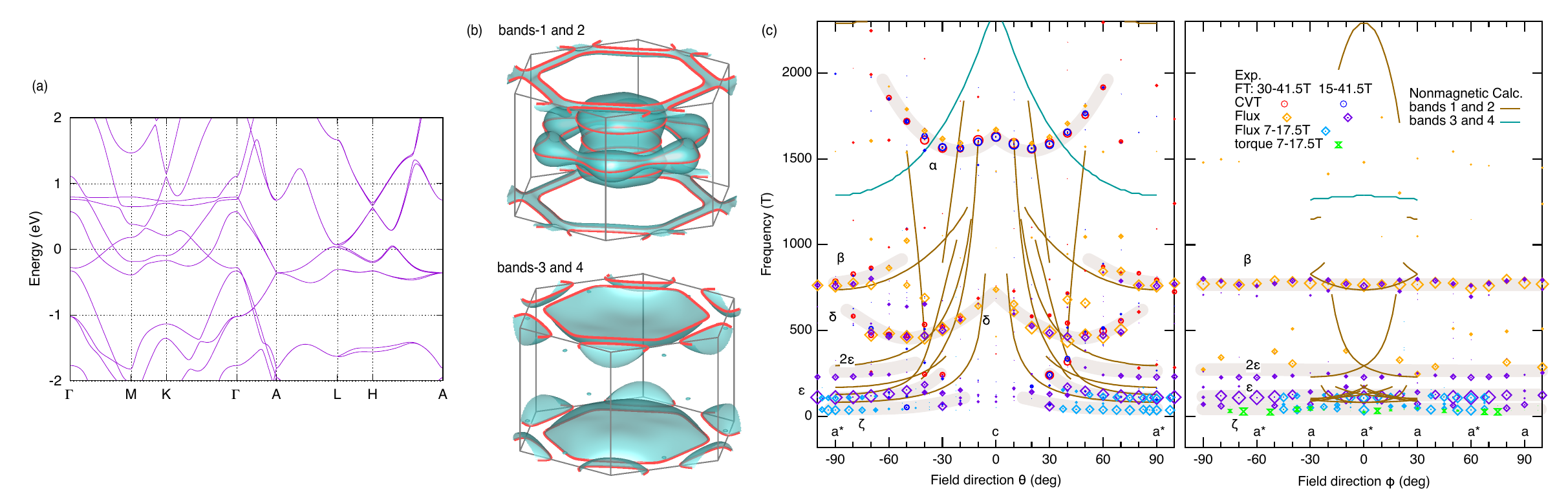}
\caption{\label{band}(a) and (b) Calculated band structure (a) and Fermi surface (b) of CrSb in the nonmagnetic state.
In (b), extremal orbits for $B \parallel c$ are indicated.
(c) Magnetic-field-direction dependences of the quantum oscillation frequencies for field rotation in the $ca^*$ and $a^*a$ planes compared to nonmagnetic calculations.
Symbol conventions are the same as in Fig. 2 of the main text, except that the solid curves plot the theoretical frequencies calculated from the nonmagnetic Fermi surface shown in (b).
}
\end{figure}

We also performed band-structure calculations for the nonmagnetic state for the sake of comparison.
The calculated nonmagnetic band structure and Fermi surface are shown in Figs. S3(a) and (b).
Figure S3(c) compares the experimental quantum-oscillation frequencies with the frequencies expected from the nonmagnetic Fermi surface.

\section{Comments on the work by Long \textit{et al}.}

\begin{figure}
\includegraphics[width=17cm]{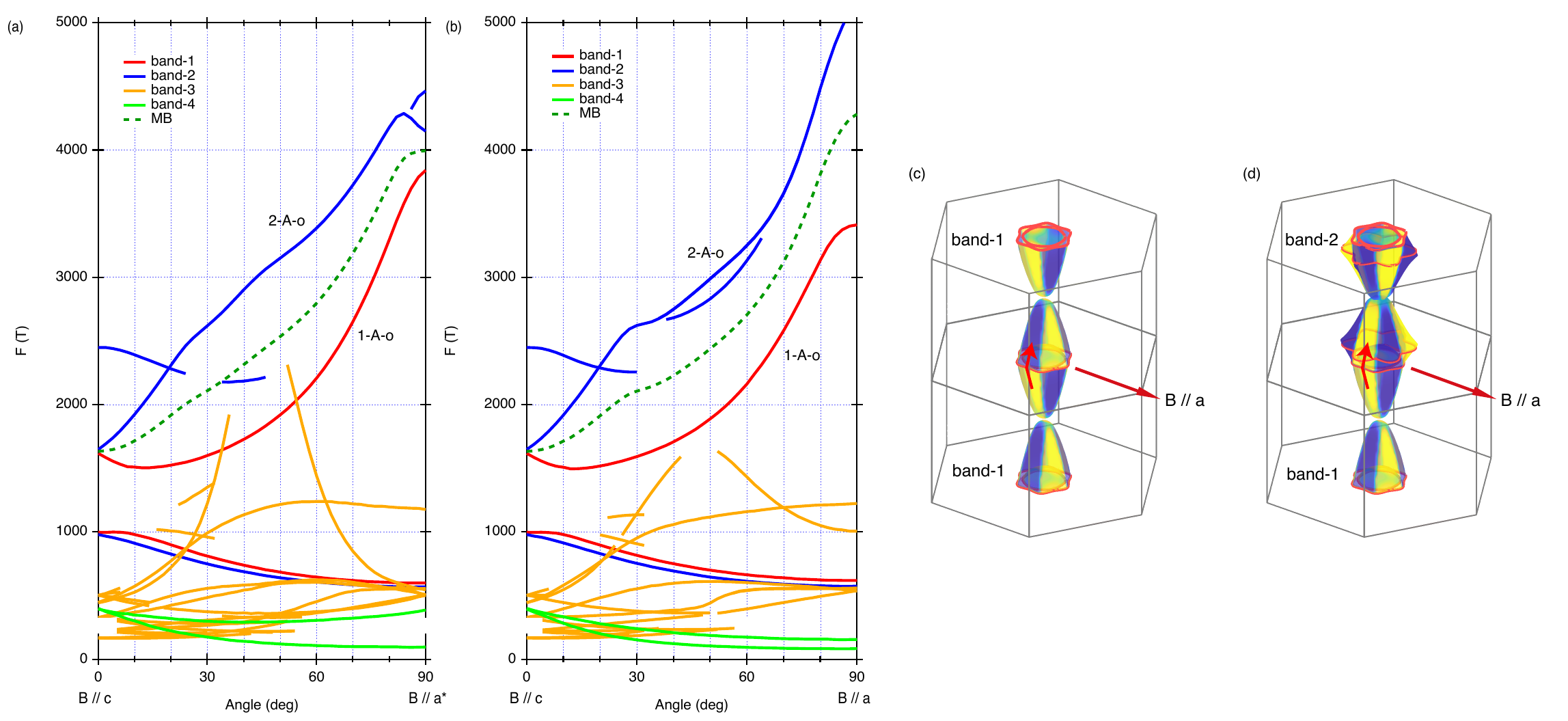}
\caption{\label{band}(a) and (b) Calculated frequencies in (a) the $ca^*$ plane and (b) the $ca$ plane for the altermagnetic Fermi surface  ($U$ = 0.03 Ry) shown in Fig. 2(c).
The green dashed lines are the magnetic breakdown frequency between 1-A-o and 2-A-o.
(c) and (d) For $B \parallel a$, when crossing the A plane, the spin orientation reverses on the 1-A-o (2-A-o) orbit (c), but it does not reverse on the MB orbit (d). 
}
\end{figure}

Figures S4(a) and (b) show the theoretical frequencies calculated from the altermagnetic Fermi surface  ($U$ = 0.03 Ry) shown Fig. 2(c) for the $ca^*$ and $ca$ planes, respectively.
These can be compared to the experimental results for the $c$-$ab$ plane and $c$-$a$ plane, respectively, by Long \textit{et al}. \cite{Long26condmat}.
The green dashed lines in the figures show the average of the frequencies 1-A-o and 2-A-o, i.e., $(F_{\mathrm{1-A-o}}+F_{\mathrm{2-A-o}})/2$, which corresponds to the magnetic breakdown (MB) frequency when MB occurs between orbits 1-A-o and 2-A-o on the A plane.

For the $ca^*$ ($c$-$ab$) plane, the frequency $f_1$ reported by Long \textit{et al.} \cite{Long26condmat} is well accounted for by the orbit 1-A-o very well, while the higher frequency $f_2$ is closer to the MB frequency than to the frequency 2-A-o.
For the $ca$ plane, the frequency $f_1$ by Long \textit{et al.} \cite{Long26condmat} is well accounted for by the orbit 1-A-o up to $\sim80^{\circ}$ from the $c$ axis, but becomes closer to the MB frequency for field angles within $\sim10^{\circ}$ of $B // a$.

At present, it is not possible to conclusively determine the observation of MB orbits; however, we remark on the following points for future consideration. 
As the A plane ($k_z=0.5$) is a nodal plane of the altermagnetic spin splitting, the splitting between bands 1 and 2 on the A plane is due to SOC and is therefore small. 
Consequently, magnetic breakdown is likely to occur.
For $B \parallel a$, when crossing the A plane, the spin orientation reverses on the 1-A-o (2-A-o) orbit, but it does not reverse on the MB orbit [Figs. S4(c) and (d)]. 
This difference in spin behavior may influence the MB probability.

\section{Landau-fan plot of $\alpha$ oscillations}
Figure S5 shows the Landau-fan plot of $\alpha$ oscillations for $B \parallel c$ based on the data shown in Fig. 1(a), assigning an integer index $n$ to each $\rho_{osc}$ peak.
A linear fitting gives a frequency of $F$ = 1627 T, very close to 1628 T determined from a Fourier transform. The $n$-axis intercept is -0.06(3), which is very close to zero, in contrast to the value of -0.1667 reported in \cite{Du26SciChinaPhys}.

The intercept cannot be straightforwardly interpreted, for the following reasons:
First, the value of $\phi_Z$ in Eq. (S1) is not readily estimated.
Second, the oscillation frequency $F_1$ from the maximum cross-section of the band-1 Fermi surface (orbit 1-A-o) approaches $F_2$ from the minimum cross-section of band-2 (orbit 2-A-o) in field directions near $B \parallel c$.
Therefore the observed $\alpha$ oscillation near $B \parallel c$ is actually a sum of both frequencies 
and the phase analysis of the $\alpha$ oscillation becomes very ambiguous.

Let us consider how the near-zero intercept that we found can be explained.
The experimentally observed oscillation $\rho_{osc}^{exp}$ is the sum of two oscillations:
\begin{equation}
\begin{split}
\rho_{osc}^{exp} = &-A_1(T, B)\cos \left[2\pi \left(\frac{F_1}{B}-\frac{1}{2}\right) - \frac{\pi}{4} + \phi_{B,1} + \phi_{Z, 1} \right] \\
&-A_2(T, B)\cos \left[2\pi \left(\frac{F_2}{B}-\frac{1}{2}\right) + \frac{\pi}{4} + \phi_{B,2} +  \phi_{Z, 2} \right].
\end{split}
\tag{S2}
\end{equation}
$F_1$ and $F_2$ are rendered slightly different by spin--orbit coupling and/or slight misorientations of the field direction.
As the two orbits are degenerate when spin--orbit coupling is absent and the magnetic field is exactly parallel to the $c$ axis, we assume $A_1 = A_2$, $\phi_{B, 1}=-\phi_{B, 2}$ and $\phi_{Z, 1} = -\phi_{Z, 2}$.
We then obtain 
\begin{equation}
\rho_{osc}^{exp} = -2A_1(T, B)\cos \left( \frac{\Delta F}{B} \pi + \frac{\pi}{4} - \phi_{B, 1} - \phi_{Z,1} \right) \cos \left[2\pi \left(\frac{F_{av}}{B}-\frac{1}{2}\right) \right], \tag{S3}
\end{equation}
with
\begin{equation}
\Delta F = F_2 - F_1 > 0 \textrm{ and } F_{av}=(F_1 + F_2 )/2.  \tag{S4}
\end{equation} 
Because the cosine prefactor can be positive or negative depending on $\Delta F$, $B$, $\phi_{B, 1}$, and $\phi_{Z, 1}$, the $n$-intercept also depends on these parameters.
If the cosine prefactor is positive within the investigated field range, the $n$-axis intercept of the Landau-fan plot is zero.
For example, the prefactor remains positive in the field range of Fig. S5 when $\Delta F$ = 5 T and $\phi_{B, 1} + \phi_{Z, 1}$ = $\pi$/2 (mod $2\pi$).
In this case, the prefactor is maximized at $B$ = 20 T and decreases with increasing $B$. 
This trend might explain the slight concavity in the Dingle plot of $\alpha$ in Fig. 4(b).
However, even if this assumption is correct, the Berry phase $\phi_{B, 1}$ cannot be uniquely determined.

Under the assumption of Eq. (S3), determination of the Berry phase requires observing multiple zero-crossing fields of the prefactor to fix the sum $\phi_{B, 1} + \phi_{Z, 1}$.
Furthermore, $\phi_{Z, 1}$ must be independently determined.

\begin{figure}
\includegraphics[width=8.6cm]{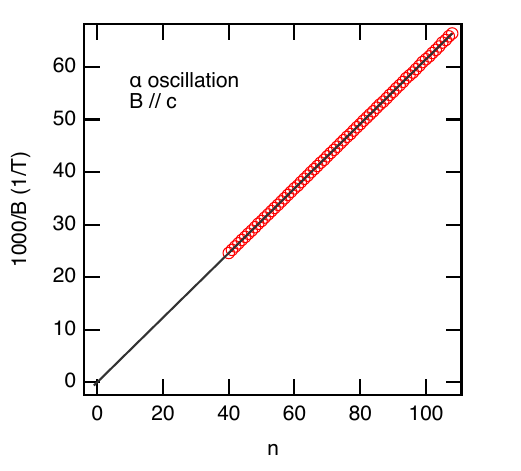}
\caption{\label{fan}Landau fan plot of $\alpha$ oscillations with $B \parallel c$.
}
\end{figure}

\section{Experimental and calculated Fermi surface parameters of C\lowercase{r}S\lowercase{b}}

\begingroup
\squeezetable
\begin{table*}
\caption{Experimental Fermi surface parameters of CrSb for $B \parallel c$, $a^*$, and $a$. 
Data for the $\delta$ frequency at $\theta = -60^{\circ}$, where the effective mass was determined, are also shown. 
The parameters for the $\alpha$ frequency are based on measurements of the CVT-grown sample, while the other parameters are based on the flux-grown sample.
The right side shows the theoretical candidate orbits, named by their band number, orbit center, and (if necessary) inner (i) or outer (o). 
}
\begin{ruledtabular}
\begin{tabular}{ccddcccdd}
\multicolumn{1}{c}{Field} & \multicolumn{5}{c}{Experiment} & \multicolumn{3}{c}{Calculation} \\
\cline{2-6} \cline{7-9}
direction & Branch & \multicolumn{1}{c}{$F$ (kT)} & \multicolumn{1}{c}{$m^*/m_e$} & \multicolumn{1}{c}{$\tau$ (ps)} & \multicolumn{1}{c}{$\mu$ (cm$^2$/(Vs)} & Orbit & \multicolumn{1}{c}{$F$ (kT)} & \multicolumn{1}{c}{$m_{band}/m_e$} \\
\hline
$B \parallel c$ & $\alpha$ & 1.63 & 1.2(1) & 0.38(4) & 560(30) & 2-A-o & 1.65 & 1.10 \\
& & & & & & 1-A-o & 1.62 & 1.16 \\
& & & & & & 1-A-i &1.00 &  0.93  \\
& & & & & & 2-A-i &0.98 &  0.86  \\
& $\delta$ & 0.74 & & & & 3-gp$_2$ &0.50 &  1.86 \\
& & & & & & 4-M &0.40 &  1.68  \\
\\

$B \parallel a^*$ & $\beta$ & 0.77 & 1.13(4) & 0.28(2) & 400(20) & 1-A-i & 0.60 &  0.31  \\
& & & & & & 2-A-i &0.57 &  0.27  \\
& & & & & & 3-gp$_1$ &0.51&  0.90  \\
& & & & & & 3-gp$_2$ &0.50 &  1.81  \\
& $\epsilon$ & 0.11 & 0.29(3) & 0.11(3) & 700(100)  & 4-M & 0.10 &  0.43  \\
& $\zeta$ & 0.04 & & & &  &  &    \\
\\

$B \parallel a$ & $\beta$ & 0.78 &  & & & 1-A-i & 0.62 &  0.46  \\
& & & & & & 2-A-i &0.57 &  0.30  \\
& & & & & & 3-gp$_1$ &0.54&  0.85  \\
 & $\epsilon$ & 0.14 &  & & & 4-M & 0.15 & 0.68   \\
 & $\zeta$ & 0.05 & 0.24(1) & & &  &  &    \\

\\
$\theta=-60^{\circ}$ & $\delta$ & 0.47 & 1.2(2) & & & 3-gp$_2$ & 0.37 & 1.40 \\

\end{tabular}
\end{ruledtabular}
\end{table*}
\endgroup

Table SI shows the experimental frequencies, effective masses, electron scattering times $\tau$, and mobilities $\mu$ for $B \parallel c$, $a^*$, and $a$, along with the theoretical candidate frequencies for comparison and the
data of the $\delta$ branch at $\theta = -60^{\circ}$.

%